\documentclass[a4paper,conference]{IEEEtran}
\usepackage{graphicx}
\usepackage{color}

\IEEEsettopmargin{t}{30mm}
\IEEEquantizetextheight{c}
\IEEEsettextwidth{14mm}{14mm}
\IEEEsetsidemargin{c}{0mm}

\usepackage[utf8]{inputenc}
\usepackage[T1]{fontenc}
\usepackage{url}
\usepackage{ifthen}
\usepackage{cite}
\usepackage[cmex10]{amsmath} 
\newcommand {\dfn} {\stackrel{\Delta} {=}}

\newcommand {\bx} {\mbox{\boldmath $x$}}
\newcommand {\by} {\mbox{\boldmath $y$}}

\newcommand {\bX} {\mbox{\boldmath $X$}}

\newcommand{\calA}{{\cal A}}
\newcommand{\calB}{{\cal B}}

\newcommand{\calH}{{\cal H}}
\newcommand{\calI}{{\cal I}}

\begin{document}
\title{On Jacob Ziv's Individual-Sequence Approach to Information Theory}

\author{%
  \IEEEauthorblockN{Neri Merhav}
  \IEEEauthorblockA{The Viterbi Faculty of ECE\\
              Technion - Israel Institute of Technology\\
              Technion City, Haifa 3200003, Israel\\
              email: merhav@technion.ac.il}
}



\maketitle

\begin{abstract}
This article stands as a tribute to the enduring legacy of Jacob Ziv and his
landmark contributions to information theory. Specifically, it delves into
the groundbreaking individual-sequence approach -- a cornerstone of Ziv's
academic pursuits. Together with Abraham Lempel, Ziv pioneered the renowned Lempel-Ziv
(LZ) algorithm, a beacon of innovation in various versions. Beyond its
original domain of universal data compression, this article underscores the
broad utility of the individual-sequence approach and the LZ algorithm across
a wide spectrum of problem areas.
As we traverse through the forthcoming pages, it will also become evident how
Ziv's visionary
approach has left an indelible mark on my own research journey, as well as
on those of numerous colleagues and former students. We shall explore, not
only the technical power of the LZ algorithm, but also its profound impact on shaping
the landscape of information theory and its applications.
\end{abstract}

\section{Introduction}

Jacob Ziv (November 27, 1931 -- March 25, 2023), who was a luminary in the field of
information theory, served as a distinguished professor at the Electrical
Engineering Department of the Technion
in Haifa, Israel. Renowned for his groundbreaking contributions to information
theory, Ziv's work has left an indelible mark on the academic landscape. His
achievements were so profound and influential that they garnered widespread
recognition and numerous prestigious awards. Among his accolades, he was
honored with the Israel Prize for Exact Sciences in 1993, the IEEE Richard 
W.~Hamming Medal in 1995 for his invaluable contributions to information theory
and data compression, and the Claude E.~Shannon Award in 1997. His innovative
spirit was further acknowledged with the Golden Jubilee Award for
Technological Innovation in 1998 and the 2008 BBVA Foundation Frontiers of
Knowledge Award in Information and Communication Technologies. In a crowning
achievement, he was bestowed with the IEEE Medal of Honor in 2021, the highest
honor from IEEE, in acknowledgment of his fundamental contributions to
information theory, data compression technology, and his exemplary research
leadership.

In this article, I chose to focus will hone in on one important facet of Jacob Ziv's
illustrious research area that stands as a testament to his ingenuity and
dedication -- the {\em individual-sequence approach}, which I have always
found elegant and fascinating. Ziv's pioneering work in this
realm spans nearly half a century, marked by relentless creation of
brilliant innovative ideas.

During the latter half of the 1970s, Jacob Ziv and Abraham Lempel introduced a
groundbreaking shift in information theory \cite{Ziv78}, \cite{ZL77},
\cite{ZL78}. Departing from the conventional
probabilistic paradigm, which characterized sources and channels with known
statistical properties, often memoryless in structure, they envisioned a new
approach, which is the individual-sequence approach combined with
finite-state (FS) encoders/decoders, offering a fresh perspective on
universal data compression techniques and on 
coded communication in general. It was within this paradigm that the seeds of the LZ
algorithm were sown, culminating in its first two versions, in 1977 and 1978
-- the LZ77 and LZ78 algorithms, respectively.

Countless words have already been dedicated in the scientific literature to the illustrious LZ
algorithms, lauded for being rare examples of possible 
coexistence of an elegant theory and remarkable practicality.
Their profound influence, together with those of later versions of the LZ
algorithm, reverberates through the fabric
of modern life, touching each and every individual who possesses a computer,
a smart-phone, or any device that stores digital information.

Less commonly recognized are the additional pillars of the
individual-sequence approach, alongside the lesser-known versatility of the
LZ algorithms, especially, the LZ78 version. Beyond its renowned role in universal data compression, the
LZ78 algorithm turns out to serves as a potent engine for an array of information
processing tasks spanning
universal channel decoding, prediction, hypothesis testing, model order
estimation, guessing, filtering, and more. Remarkably, the asymptotic optimality
of the LZ78 algorithm as a data compressor induces its asymptotic optimality in
all these tasks as well.

This article delves into this facet of the LZ
algorithm, a subject that has always captivated my interest immensely. As we traverse
through the annals of previous research in this domain, I will not only
highlight the contributions of Ziv and his collaborators, but also shed light
on the works of other researchers who have been inspired by the
individual-sequence approach. Among them, I will draw from my own experiences,
as well as those of esteemed colleagues and former Ph.D.\ students.

\section{Modeling Approaches and Sequence Complexity}

Traditionally, since the days of Shannon, information theory has been grounded
in probabilistic models, particularly focusing on memoryless sources and
channels. Also, classical coding theorems operate under the assumption that both the
encoder and decoder have full knowledge of these sources and channels. While
these two
assumptions -- the assumption of a memoryless structure, and the assumption
that the source/channel is known, are not necessarily reflective of
reality, they persisted because they serve for an excellent simplification.
This simplification greatly facilitates the analysis and the derivation of
non-trivial bounds, offering valuable insights and understanding. Importantly,
many of these insights extend beyond the scope of known memoryless sources and
channels.

Soon after the inception of information theory, we observed the emergence of
research endeavors aimed at relaxing these two fundamental assumptions.
Departing from the memorylessness assumption led to expansions of source
coding theorems, encompassing models such as Markov sources, unifilar
finite-state sources, hidden Markov sources, and more general stationary and
ergodic sources. Similar strides were made in the realm of channel coding
theorems and their corresponding channel models.

Regarding the perspective of discarding the assumption of known statistics,
two main avenues of research have emerged. The first draws from the field of
{\em robust statistics}, wherein the approach entails assuming that the actual
source (or channel) lies within a certain neighborhood of a known nominal
model. Designs are then crafted to address the worst-case scenario within this
neighborhood. This has spurred the development of robust hypothesis testing,
particularly robust detection, robust parameter estimation, robust filtering,
and robust signal processing in general. The second route is associated with
the advancement of {\em universal methods}, which are sub-optimal schemes that
asymptotically achieve optimality in the limit of large amounts of data or large
blocks, as they adapt to the underlying source statistics. 
Certainly, within the realm of source coding, we have witnessed a progressive
evolution towards devising universal schemes capable of accommodating
increasingly diverse classes of sources (at the price of a slow-down in the
convergence towards to the entropy rate). This evolution commenced with the
treatment of the class of memoryless sources, then extended to encompass classes of Markov
and finite-state sources, culminating in non-parametric classes such as all
stationary and ergodic sources with a finite alphabet. Furthermore, atop these
advancements lies the {\em individual-sequence approach}, which treats the source
data as a deterministic entity devoid of any underlying probabilistic
mechanism. Fig.\ 1 illustrates this hierarchy of stages of departure from the
assumption of known statistics and gradually increasing the degree of
generality.

Alongside the development of universal data compression schemes, the concept
of {\em complexity}, a.k.a.\ {\em compressibility}, has emerged. While in
traditional probabilistic settings, complexity is naturally measured by the
entropy rate of the source, the individual-sequence setting presents a
challenge. Here, defining complexity is not straightforward because without
constraints on compression and decompression resources, there exists no
non-trivial lower bound on achievable compression ratios for individual
sequences. Consider, for example, an ``encoder'' that represents a given individual sequence
with a single bit, say '0', while all other possible sequences are represented by
the flag-bit '1' followed by a copy of the uncompressed input. In this scenario, the
compression ratio for the given sequence approaches zero, rendering the issue
trivial, uninteresting, and essentially useless for anything beyond that
specific sequence. This echoes the effect of overfitting in model learning,
where an overly complex model fails to generalize.
A natural expectation from a reasonable definition of complexity is that it should
converge to the entropy rate when applied to typical sequences drawn from a
random process.

One of the most famous pioneers in the
context of complexity of an individual sequence was Kolmogorov
\cite{Kolmogorov65}, \cite{Kolmogorov68}, who during the 1960s, took the algorithmic approach
and defined complexity in terms of the length of the shortest computer
program, running on a universal Turing machine,
that generates the given sequence (see also
\cite[Chap.\ 14]{CT06}). The ideas of Kolmogorov were raised independently and
nearly at the same time also by Solomonoff \cite{Solomonoff64} and Chaitin
\cite{Chaitin66}. While immensely powerful and elegant, the Kolmogorov complexity
suffers a significant limitation: it is not computable, making it challenging to
practically utilize.

About a decade later, during the latter half the 1970s, Ziv and Lempel published
a series of landmark papers \cite{LZ76}, \cite{Ziv78}, \cite{ZL77},
\cite{ZL78}, that have ultimately laid the foundation for the development their
individual-sequence approach along with their definition of sequence
complexity, termed the {\em finite-state complexity}, or the
{\em finite-state compressibility}. The finite-state compressibility of an
infinite sequence means the best achievable compression ratio that can be
achieved by any information lossless finite-state encoder, where the limit
on the number of states, $s$, that grows without bound is taken after the limit of
the length, $n$, of the sequence is taken to infinity, namely, an asymptotic
regime where $s \ll n$, to meet practicality considerations and to avoid the `overfitting' problem described earlier.

The finite-state complexity measure is not as powerful as the
Kolmogorov complexity. As an extreme (but simple) example, consider the counting sequence,
$$0100011011000001010011100101110111...$$
which is formed as a concatenation of all binary strings of length 1, followed by all
binary strings of length 2 (in lexigraphical order), and so on. Indeed, this
can be seen by parsing this sequence as follows:
$$0,1,00,01,10,11,000,001,010,011,100,101,110,111...$$
This description of the rule behind the sequence generation can
easily be translated into a very short and simple computer program, which suggests that
the Kolmogorov complexity, normalized by the sequence length $n$, tends to
zero, and so, the Kolmogorov complexity of the infinite counting sequence is
zero. On the other hand, as shown in \cite{ZL78}, the finite-state complexity
of the counting sequence is $1$, which means that this sequence is not
compressible by any information lossless finite-state encoder. 

On the bright side, the finite-state complexity is computable in contrast
to the Kolmogorov complexity, and it also satisfies the above mentioned
desired property of convergence to the entropy rate when the sequence emerges
from a stationary and ergodic source \cite{ZL78}. In the next section, we will
have more to say about it as well as and on the LZ78 algorithm and other versions of
the LZ algorithm.

\begin{figure}[h]
\hspace*{1cm}\input{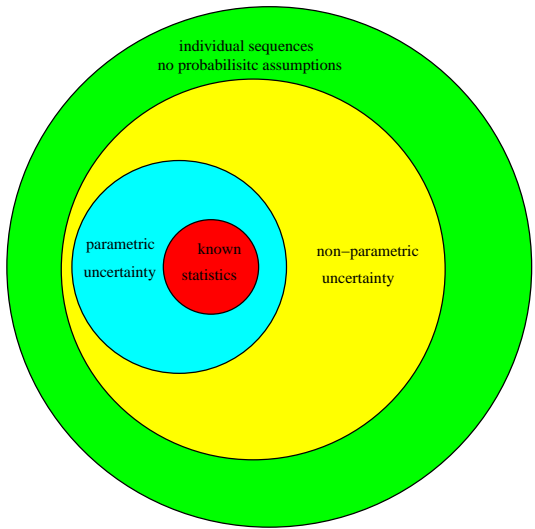}
\caption{The hierarchy of classes of sources with various degrees of
generality.}
\end{figure}

\section{Compression of Individual Sequences by FSM's - the LZ Algorithm}

Following Ziv and Lempel \cite{ZL78},
consider the system depicted in Fig.\ 2, which describes a source sequence
fed into a finite-state encoder for the purpose of
lossless data compression, and outputs the compressed representation. More precisely, the encoding mechanism is as follows:
An infinite individual source sequence, $\bx=(x_1,x_2,\ldots)$, from a finite
alphabet serves as an input to a finite-state encoder that implements
recursively the following two equations, for $i=1,2,\ldots$:
\begin{equation}
\label{output}
y_i=f(z_i,x_i),
\end{equation}
and
\begin{equation}
\label{next-state}
z_{i+1}=g(z_i,x_i),
\end{equation}
where $y_i$, the encoder output at time instant $i$, is a variable-length
binary string, whose length, $\ell(y_i)$, may sometimes be zero (no output), when the encoder idles, and $z_i$
is the encoder state, which takes on values in a finite set of states of size
$s$. Generally speaking, the state represents whatever the encoder ``remembers''
from the past of the input, for example, the state could be defined by a shift
register, $z_i=(x_{i-1},x_{i-2},\ldots,x_{i-k})$, that stores the $k$ most
recent source inputs, if $g$ is chosen accordingly. Also, it is assumed that
the encoder is information lossless, which means that the source can be
reconstructed from any segment of the compressed output, provided that the
states at the beginning and at the end of this segment are provided as well.
The finite-state compressibility is then defined in several steps. First,
define
\begin{equation}
\label{1ststep}
\rho_s(x_1,\ldots,x_n)=\min_{\{\mbox{$s$-state
encoders}\}}\frac{\sum_{i=1}^n\ell(y_i)}{n},
\end{equation}
which is the best compression ratio that can be attained among all $s$-state
encoders, $(f,g)$. Next, define
\begin{equation}
\rho_s(\bx)=\limsup_{n\to\infty} \rho_s(x_1,\ldots,x_n),
\end{equation}
and finally, define the {\em finite-state compressibility} of $\bx$ by
\begin{equation}
\rho(\bx)=\lim_{s\to\infty}\rho_s(\bx).
\end{equation}

\begin{figure}[h]
\hspace*{0.5cm}\input{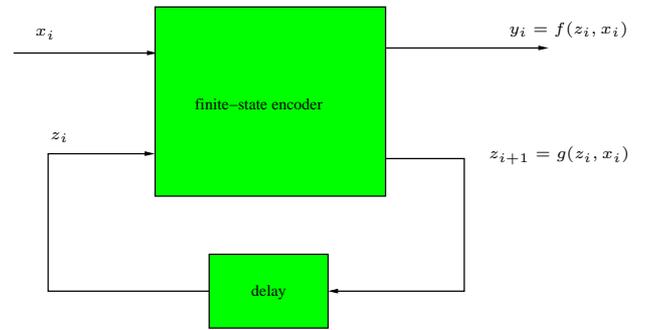}
\caption{Finite-state encoder.}
\end{figure}

While the sequence of minimizing encoders in (\ref{1ststep}) depends on
$(x_1,\ldots,x_n)$ for $n=1,2,\ldots$, and a given $s$, the quest is for a
{\em single} encoder that asymptotically attains $\rho(\bx)$ for every $\bx$,
namely, a universal data compression scheme in the individual-sequence sense.
The LZ78 algorithm, proposed in \cite{ZL78} and briefly described next, achieves this objective.

The main engine of the LZ78 algorithm is the so called {\em incremental
parsing procedure}, which is a sequential process of parsing the source
string to distinct phrases, where each phrase is the shortest substring that
has not been observed before as parsed phrase, and the last phrase might be
incomplete. One example was already shown
above in the context of the counting sequence. As another example, the
string:\\

\vspace{0.05cm}

\noindent
{\tt repeatandrepeatandrepeatandrepeatandrepeat}\\

\noindent
is parsed as:\\

\vspace{0.05cm}

\noindent
{\tt r,e,p,ea,t,a,n,d,re,pe,at,an,dr,ep,eat,}\\
{\tt and,rep,eata,nd,repe,at}\\

Let $n$ denote the length of source string, $x_1,x_2,\ldots,x_n$, and let $c$
denote the number of parsed phrases. In the above example, $n=42$ and $c=21$.
Clearly, when the source string exhibits a high degree of repetitive behavior,
the phrases grow rapidly along the process and then their number, $c$, is
relatively small for a given string length, $n$. Conversely, if the string has
a low level of repetitiveness, the phrases grow slowly as we proceed in the
parsing process, and then $c$ is very large. It is therefore plausible that
$c$, or any monotonically increasing function of $c$, may serve as a measure
of the complexity of the source string. As shown in \cite{ZL78}, it turns out that the relevant
measure of complexity, as far as data compression is concerned, is essentially
given by the function $c\log c$, or actually, $\frac{c\log c}{n}$, after normalizing by
$n$, in order to give it the meaning of a compression ratio. 

Indeed, the main
results in \cite{ZL78} are given by a coding theorem and its converse in that
respect: On the one hand, the converse theorem asserts that if $s \ll n$, 
then $\rho_s(x_1,\ldots,x_n)$ cannot be much smaller than the LZ complexity,
defined as
\begin{equation}
\rho_{\mbox{\tiny LZ}}(x^n)\dfn\frac{c\log c}{n}.
\end{equation}
On the other hand, the coding theorem tells that 
$\rho_{\mbox{\tiny LZ}}(x^n)$ is an
essentially achievable compression ratio (up to a vanishingly small redundancy
term), and the proof of the latter theorem is constructive -- by performance analysis of the LZ78
algorithm, which, roughly speaking works as follows:
\begin{enumerate}
\item Apply the incremental parsing procedure to the source string, $(x_1,\ldots,x_n)$.
\item Compress each parsed phrase sequentially as follows:
\begin{enumerate}
\item Letting $l$ denote the length of the current phrase, compress the
substring formed by the first
$l-1$ symbols by indicating the location of an earlier (already decoded) phrase of length $l-1$
with matching contents.
\item Encode the last symbol of the current phrase but its binary representation,
without compression.
\end{enumerate}
\end{enumerate}

There is a certain caveat, however, in the sense that this coding theorem and its
converse are not quite compatible with each other, because the number of
states needed to implement the LZ algorithm over a source block of length $n$
is not negligible compared to $n$ as it should be according to the converse
theorem. On the contrary, it even grows exponentially with $n$, because the
entire block should be stored at the encoder in order to implement it. 
This incongruity between the coding theorem and the converse theorem is closed 
once the limit of $s\to\infty$ is taken. But this limit should be
taken cautiously. Specifically, if one
restarts the LZ algorithm for every block of length, say $k$ (in order to
limit the number of states), and considers the quantity,
$$\limsup_{k\to\infty}\limsup_{n\to\infty}\frac{k}{n}\sum_{i=0}^{n/k-1}
\rho_{\mbox{\tiny LZ}}(x_{ik+1},x_{ik+2},\ldots,x_{ik+k}),$$
which achieves $\rho(\bx)$ in the limit of $s\to\infty$, then the gap is
indeed closed.

A simplistic point of view on the quantity $c\log c$ could be the following:
Consider the $c$
distinct phrases as super-letters over a super-alphabet (or dictionary) of variable length
strings, each of with appears in $(x_1,\ldots,x_n)$ exactly once, and so,
their empirical probabilities are all equal to $1/c$. Accordingly, ignoring
integer length constraints, the
code-length to be assigned to each such phrase is $-\log(1/c)=\log c$. Since
we have a total of $c$ phrases to compress, and each one is represented by $\log c$ bits, the
total length is $c\log c$. 
This perspective, however, is overly simplistic because the decoder lacks
explicit foreknowledge of the contents of these super-letters. Interestingly,
the LZ78 algorithm achieves a compression ratio of approximately $c\log c$
even without necessitating an explicit header to inform the decoder about the
phrase contents.

The LZ complexity, $\rho_{\mbox{\tiny LZ}}(x_1,\ldots,x_n)$, can be thought of
as the individual-sequence analogue of the entropy in the sense that when
$(x_1,\ldots,x_n)$ is a typical realization of a stationary and ergodic
source, $\rho_{\mbox{\tiny LZ}}(x_1,\ldots,x_n)$ converges to the entropy rate
of that source. More precisely, if $\bX=(X_1,X_2,\ldots)$
is a stationary and ergodic source, then $\{\rho_{\mbox{\tiny
LZ}}(X_1,\ldots,X_n),~n=1,2,\ldots\}$ converges to the entropy rate almost
surely \cite{ZL78}. In other words, the LZ78 algorithm is universal for all
stationary and ergodic sources in quite a strong sense (almost sure
convergence and not just expectation).

The LZ78 algorithm is only one among an array of quite many versions of
the LZ algorithm. The common feature of all of those versions is that they take
advantage of repetitiveness in the source sequence to be compressed, by applying various mechanisms of {\em string
matching}. As another example, the LZ77 algorithm \cite{ZL77} is based on
storing a large sliding window of the most recent past symbols observed and seeking
the longest match that can be found within the window for the current string
being compressed. Compression is obtained by encoding two positive
integers: the length of the matching string and the shift needed to point on its most
recent earlier occurrence.

The impact of LZ algorithms is indeed profound, representing some of the most
widely employed techniques for lossless data compression. Among these, DEFLATE
stands out as a variant tailored for optimizing decompression speed and
compression ratio. Notably, in the 1980s, spurred by the work of T.~Welch, the
Lempel-Ziv-Welch (LZW) algorithm emerged as the preferred method for a wide
array of compression applications. Its versatility is evident in its adoption
across various domains: from GIF images and compression utilities like PKZIP
to hardware peripherals such as modems. Moreover, it underpins the compression
of file formats like PDF, TIFF, PNG, ZIP, as well as popular video formats
like MP3, and finds utility in cell phones.
Remarkably, the ubiquity of LZ compression extends to everyday devices such as
desktop computers, laptops, and smart-phones, where it quietly operates in the
background, seamlessly managing digital information storage. It is a testament
to the algorithm's efficiency that countless individuals interact with LZ
compression on a daily basis without necessarily being aware of its presence.
Given its monumental significance, it is no wonder that in 2004, the IEEE
recognized the LZ algorithm as a Milestone in Electrical Engineering and
Computing, solidifying its place in the annals of technological advancement.

Earlier, it was mentioned that the LZ complexity can be viewed as the individual-sequence
analogue of the entropy rate. On the other hand, it is well known that the concept of
entropy is fundamental, not only in information theory, but also in
thermodynamics and statistical physics. First and foremost, it plays the
central role in the second law of
thermodynamics which asserts that the total entropy of an isolated system
cannot decrease.
It turns out then that the LZ
complexity may play the role of the individual-sequence analogue of entropy
also in statistical physics. Following the 
thought provoking ideas behind the famous
Maxwell demon and the Szilard engine,
a recent research trend in statistical
physics has been evolving around physical systems that, in addition to the
traditional heat
reservoir at fixed temperature, include also an information reservoir in the
form of a digital memory device or a magnetic tape with random digital information
stored on it, like the one depicted in Fig.\ 3 (see, e.g., \cite{me15} and
references therein). The main theme in these works
is in extending the second law of thermodynamics in a way that includes also
a term pertaining to the change in the entropy of the information reservoir. 
Another way to look at the extended second law is to observe that it is
possible to convert heat emanating from the heat reservoir (not shown in Fig.\ 3)
to mechanical work at the cost of
writing information in the information reservoir, that is, increasing its
entropy. In Fig.\ 3, 
we describe a system with a certain mechanism of sequential interaction between a digital tape and
a simple mechanical system from which one can extract work in the form of
lifting the mass, $m$. As shown in \cite{me15}, if the information recorded on
the tape is an individual sequence of bits, rather than random data, then the
same extended version of the second law continues to hold with the Shannon
entropy being replaced by the LZ complexity. More details can be found in
\cite{me15}.
 
\begin{figure}[h]
\hspace*{0.5cm}\input{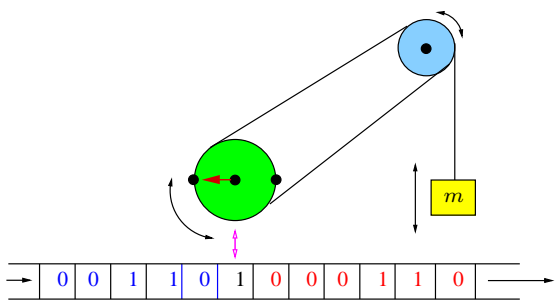}
\caption{Physical system with an information reservoir}
\end{figure}

The paradigm of compression of individual sequences using finite-state encoders/decoders have
has been extended by Ziv and by others, in several directions, including the presence of side
information, settings of distributed coding, and lossy compression with and without side
information, see, e.g., \cite{me14}, \cite{me21}, \cite{MZ06}, \cite{UK03},
\cite{Ziv80}, \cite{Ziv84} and \cite{Ziv85}, for a non-exhaustive sample of references.

However, instead of reviewing all these extensions, I believe it would be more
interesting to devote the last section of this article to another aspect of
Ziv's work, which is the utility of the LZ78 algorithm, or more precisely -- the
incremental parsing procedure associated with it, in a large variety of
information processing tasks beyond compression. In particular, it is
fascinating that the asymptotic optimality property of the LZ78 algorithm (in the
compression sense) is `inherited' when it is utilized in those other tasks, 
resulting in asymptotically optimal schemes in each and every one of them. 
This indicates that there must be something very deep and powerful in the incremental parsing procedure 
for the purpose of gathering statistics in a very general sense, that includes
even individual sequences.

\section{The LZ Algorithm at the Service of Tasks Beyond Compression}

One of the pivotal tools for deriving and
developing many of the results in the context of ``the LZ algorithm for tasks beyond
compression'' is known as {\em Ziv's
inequality} \cite[Lemma 13.5.5]{CT06}, \cite{PWZ92}, which asserts that the
probability of any string, $(x_1,\ldots,x_n)$, under any Markov source of any
order, or any
general finite-state source, or even a hidden Markov source, cannot be larger
than $2^{-c\log c}$ up-to a possible factor that grows in a sub-exponential
rate as a function of $n$, or equivalently,
\begin{equation}
\log P(x_1,\ldots,x_n)\le -c\log c +n\epsilon_n,
\end{equation}
where $\epsilon_n$ tends to zero as $n$ tends to infinity.
This inequality is interesting also on its own right.

At first glance, it might seem intriguing that
there is any connection whatsoever between the probability of a sequence and
the number of phrases, $c$. This relationship stems from a combinatorial
consideration of lower bounding the number of sequences of length $n$ that
share the same probability as $(x_1,\ldots,x_n)$, by counting phrase permutations
and showing that their number is exponentially lower bounded by $2^{c\log c}$,
thus echoing parallel well known results from the method of types.

We next review briefly some applications of the LZ algorithm in several
problem areas, other than data compression. It should be pointed out that this
is by no means the full set of applications.

\subsection{Hypothesis Testing and Model Order Estimation}

About a decade after the invention of the LZ algorithm, Ziv considered a
certain class of
problems of universal hypothesis testing \cite{Ziv88a}, \cite{Ziv88b}. The
simplest problem in this class is the following: Given a binary sequence,
$(x_1,\ldots,x_n)$, which is a realization of a certain random process,
the task is to decide between two hypotheses:\\

\noindent
$\calH_0$: $x_1,\ldots,x_n$ are independent fair coin tosses.\\
$\calH_1$: $x_1,\ldots,x_n$ are {\em not} independent fair coin tosses.\\

One motivation for this problem could be testing the reliability of a random
number generator for the purpose of simulations.

While under $\calH_0$, the probability of $(x_1,\ldots,x_n)$
is given simply by $P_0(x_1,\ldots,x_n)=2^{-n}$, the difficulty is that under
$\calH_1$, we know nothing about the underlying probability distribution,
except that it is not a binary symmetric source, and so, it is impossible to
apply the optimal likelihood ratio test (LRT).

Nonetheless, adopting the Neymann-Pearson criterion for binary hypothesis
testing, consider a class of discriminators that are implementable by finite
state machines with $s$ states. Such a finite state machine recursively implements a
next-state function,
\begin{equation}
\label{nextstate}
z_{i+1}=g(z_i,x_i),~~~~i=1,2,\ldots,n,
\end{equation}
and stores the matrix of all joint counts, 
\begin{equation}
n(x,z)=\sum_{i=1}^n\calI\{x_i=x,z_i=z\}
\end{equation}
for all possible combinations of $(x,z)$. A decision rule is then a partition
of the space of space of matrices into two regions, $\calA_0$ and $\calA_1$,
where in $\calA_i$ one makes the decision in favor of $\calH_i$, $i=0,1$.
The motivation for considering such a structure is that it includes the optimal LRT
as a special case
whenever the source $P_1$, under $\calH_1$, is a finite-state source,
characterized by the product form,
\begin{equation}
P_1(x_1,\ldots,x_n)=\prod_{i=1}^nQ(x_i|z_i),
\end{equation}
with $\{z_i\}$ being generated by (\ref{nextstate}).

Consider the decision rule,
\begin{equation}
\label{dr1}
\mbox{decision}=\left\{\begin{array}{ll}
\calH_0 &\rho_{\mbox{\tiny LZ}}(x_1,\ldots,x_n)\ge 1-\lambda\\
\calH_1 &\rho_{\mbox{\tiny LZ}}(x_1,\ldots,x_n)< 1-\lambda,\end{array}\right.
\end{equation}
where $0< \lambda < 1$ is a prescribed constant.

It turns out that this decision rule uniformly minimizes the probability of error given
$\calH_1$ among all decision rules of the above described structure (for any
$g$ and any partition) among all decision rules for which the probability of
error given $\calH_0$ decays exponentially at least as fast as
$2^{-\lambda n}$.

This simple decision rule is intuitively appealing: to determine whether
$x_1,\ldots,x_n$ are fair coin tosses or not, let us compress it by the LZ78
algorithm and compare the resulting compression ratio to a threshold, as a
sequence of independent fair coin tosses is incompressible. The choice of
$\lambda$ depends on the false-alarm probability that we are willing to
accept,

More generally, suppose that under $H_0$,
the underlying process
is memoryless, but otherwise unknown. 
In other words, we are supposed to decide whether $(x_1,\ldots,x_n)$
emerges from some memoryless source or not.
Then, the corresponding extension of
(\ref{dr1}) is in replacing the term $1$ by the (memoryless) empirical entropy, $\hat{H}$, of
$(x_1,\ldots,x_n)$, or equivalently, 
\begin{equation}
\label{dr2}
\mbox{decision}=\left\{\begin{array}{ll}
\calH_0 &\hat{H}-\rho_{\mbox{\tiny LZ}}(x_1,\ldots,x_n)\le \lambda\\
\calH_1 &\hat{H}-\rho_{\mbox{\tiny LZ}}(x_1,\ldots,x_n)>
\lambda.\end{array}\right.
\end{equation}
Here the intuition is that we compare the code-lengths of two universal data compression schemes,
one designed for the class of memoryless sources, whose compression ratio is
about $\hat{H}$, and the other is the LZ algorithm which is far more general.
If the difference is below some threshold, consider the source to be
memoryless.

In \cite{MGZ89}, this idea was further generalized to the problem of estimation of the order of
a Markov source with an asymptotically optimal trade-off between the
underestimation and the overestimation probabilities. Denoting by $\hat{H}_k$
the empirical entropy of $(x_1,\ldots,x_n)$ under $k$-th order Markov modeling,
the order estimator,
\begin{equation}
\hat{k}=\min\{k:~\hat{H}_k-\rho_{\mbox{\tiny LZ}}(x_1,\ldots,x_n)\le\lambda\}
\end{equation}
turns out to uniformly minimize the underestimation probability among all
model order estimators for which the overestimation probability decays at an
exponential rate at least as fast as $2^{-\lambda n}$. 

Yet another
generalization of this line of work, which is about estimating the number of states
of a non-unifilar finite-state source (a.k.a.\ hidden Markov source), can be
found in \cite{ZM92}. Additional results concerning tests for randomness and
tests for independence can be found in \cite{Ziv88b}.

\subsection{A Measure of Divergence between Sequences} 

Can we tell when two individual sequences are ``statistically similar'' and
when they are not? Intuitively, we feel that two sequences like
$$00001000010000001~~\mbox{and}~100001000000010000$$
do have ``very similar statistical characteristics'' whereas two sequences
such as
$$00001000010000001~~\mbox{and}~111101111001111010$$
do not. What could be a good measure of statistical resemblance between two
individual sequences in general, whatever the meaning of such a term might be?

In \cite{ZM93}, an attempt was made to define a certain metric that quantifies
the statistical similarity/dissimilarity between two individual sequences,
with application to universal classification using training data.
Specifically for two finite-alphabet individual sequences, $x^n=(x_1,\ldots,x_n)$ and
$y^n=(y_1,\ldots,y_n)$, let:
\begin{equation}
\Delta(x^n\|y^n)=\frac{c(x^n\leftarrow y^n)\log n-c(x^n)\log c(x^n)}{n}
\end{equation}
where
$c(x^n)$ is $c$ as before and
$c(x^n\leftarrow y^n)$ is the number of phrases of $x^n$
with respect to $\by$, created in the following manner:\\
\begin{enumerate}
\item Find the longest prefix string of $x^n$ that appears somewhere in $y^n$,
namely, the
largest $i$ such that $(x_1,x_2,\ldots,x_i)=(y_j,y_{j+1},\ldots,y_{j+i-1})$
for some $j$.
\item Continue from $x_{i+1}$ in the same manner until $x^n$ is exhausted.
\end{enumerate}

If $x^n$ and $y^n$ are `statistically similar', the phrases of $x^n$ w.r.t.\
$y^n$ are long
and then $c(x^n\leftarrow y^n)$ is relatively small, which implies small
$\Delta(x^n\|y^n)$.
As an example, let $n=11$, $x^{11}=(01111000110)$ and $y^{11}=(10010100110)$.
Parsing $x^{11}$ with respect to $y^{11}$ yields $(011,110,00110)$, and so,
$c(x^{11}\leftarrow y^{11})=3$.

While the LZ complexity is the individual-sequence analogue of the entropy
rate, it turns out that $\Delta(x^n\|y^n)$ is the individual-sequence analogue
to the relative entropy (or the Kullback-Leibler divergence) between two
probability distributions. In \cite{ZM93}, it is shown that
$\Delta(x^n\|y^n)$ can be used for universal classification using training
data. In particular, it discriminates between statistically distinguishable sequences
whenever there is some finite-state classifier that can do this task.

This individual-sequence divergence between two sequences finds its
applications in several disciplines, such as text classification \cite{PCF05},
ECG-based personal identification and authentication \cite{PCFF10},
anomaly detection \cite{Oh23}, and also for divergence estimation
in the context of assessing
entropy production and energy dissipation in certain processes that take place
in physical systems out of equilibrium \cite{GRSPMLAC21}.

\subsection{Universal Channel Decoding}

In 1985, Ziv proposed a universal decoder for unifilar finite-state channels,
which achieves the same random coding error exponent as that of the optimal maximum
likelihood (ML) decoder with respect to the ensemble of random codebooks, whose
codewords are drawn independently under the uniform distribution \cite{Ziv85}. 

By ``unifilar finite-state channel'', we mean a channel that admits the
product-form strructure
\begin{equation}
P(y^n|x^n)=\prod_{i=1}^nP(y_i|x_i,z_i),
\end{equation}
where $z_i$ is the channel state at time $i$, which obeys the recursion,
\begin{equation}
z_{i+1}=q(x_i,y_i,z_i), ~~~~~~i=1,2,\ldots,n,
\end{equation}
for some next-state function $q$.

Ziv's universal decoding metric is defined as follows. Let
$x^n=(x_1,\ldots,x_n)$ be a channel
input vector (a codeword) and let
$y^n=(y_1,\ldots,y_n)$ be a channel output vector. 
Define $c(x^n,y^n)$ to be the number of phrases in joint parsing of
$((x_1,y_1),(x_2,y_2),\ldots(x_n,y_n))$, and denote by
$c(y^n)$ the number of distinct phrases of $y^n$. Finally, let
$c_\ell(x^n|y^n)$ be number of repetitions of the $\ell$th distinct
phrase of $y^n$, which is equal to the number of distinct phrases of $x^n$
that are aligned to the $\ell$th distinct phrase of $y^n$, $1\le\ell\le
c(y^n)$. For example,\footnote{The same example appears also in \cite{Ziv85}.} let $n=6$ and 
\begin{equation}
\left(\begin{array}{c}
x^6\\
y^6\end{array}\right)=\left(\begin{array}{ccccccccc}
0 & \bigg| & 1 & \bigg| & 0 & 0 & \bigg| & 0 & 1\\
0 & \bigg| & 1 & \bigg| & 0 & 1 & \bigg| & 0 & 1\end{array}\right).
\end{equation}
Then, $c(y^6)=3$ and
\begin{equation}
c_1(x^6|y^6)=
c_2(x^6|y^6)=1;~~c_3(x^6|y^6)=2.
\end{equation}
The universal decoding metric is defined as
\begin{equation}
u(x^n|y^n)=\sum_{\ell=1}^{c(y^n)}c_\ell(x^n|y^n)\log c_\ell(x^n|y^n),
\end{equation}
and the proposed universal decoder selects the codeword $x^n$ with the smallest
$u(x^n,y^n)$ for the given $y^n$. 

The quantity $u(x^n|y^n)/n$
is an individual-sequence counterpart of the conditional entropy, and so,
Ziv's universal decoder echoes the well-known minimum conditional entropy
decoder, which is universal for memoryless channels. Indeed, as a byproduct of
\cite{Ziv85}, $u(x^n|y^n)/n$ is established as the conditional version of the
LZ complexity in the sense that it admits both a coding theorem and a converse
for encoding an individual sequence $x^n$ in the presence of a side
information sequence $y^n$ (available at both ends) using finite-state
encoders (see also \cite{me00}, \cite{UK03}).

In \cite{LZ98} Lapidoth and Ziv have extended the findings of \cite{Ziv85} to
non-unifilar finite-state channels, namely, channels for which the next-state
function $q$ is stochastic.

\subsection{Encryption}

In an unpublished memorandum,
\cite{Zivxx}, Ziv considered the problem of perfectly secure encryption of
individual sequences, where the eavesdropper is equipped with a finite-state
machine. More specifically, it was postulated in \cite{Zivxx} that the
eavesdropper has some
prior knowledge about the plain-text, which can be represented in terms of the
existence of some set of
``acceptable messages'' that constitutes the a-priori level of uncertainty (or
equivocation) that the
eavesdropper has concerning the source input -- the larger the acceptance
set, the larger is the
uncertainty. It was assumed that there exists an finite-state machine that can test
whether or not a given candidate
plain-text message is acceptable. If the finite-state machine produces the
all-zero sequence in
response to that message, then this message is considered acceptable. Perfect security is
then defined as a
situation where the size of the acceptance set is not reduced (and hence
neither is the uncertainty)
in the presence of the cryptogram. The main result in \cite{Zivxx} is that the
asymptotic key rate needed
for perfectly secure encryption in that sense, cannot be smaller (up to
asymptotically vanishing
terms) than the LZ complexity of the plain-text source. This
lower bound is
clearly asymptotically achieved by one-time pad encryption of the bit-stream
obtained by LZ
compression of the plain-text source. This is in perfect analogy to Shannon’s
classical probabilistic
counterpart result, asserting that the minimum required key rate is equal to
the entropy rate of
the source.

In \cite{me13} encryption of individual sequences is considered as well, but
the modeling approach
and the definition of perfect secrecy are substantially different. Rather than
assuming that the
encrypter and decrypter have unlimited resources, and that it is the
eavesdropper which has limited
resources, modeled in terms of finite-state machines, in \cite{me13}, the
opposite is true. The model adopted therein is of 
a finite-state encrypter, which receives as inputs the plain-text sequence and
the secret key bit-stream, and it produces a cipher-text. 
Accordingly, a notion of {\em finite-state encryptability} is defined as the
minimum achievable rate
at which key bits must be consumed by any finite-state encrypter in order to
guarantee perfect
security against an unauthorized party in the sense that the probability
distribution of the cryptogram is independent of the plaintext input.
The final conclusion in \cite{me13} is the same as in \cite{Zivxx}: the finite-state
encryptability is equal
to the finite-state compressibility, which in turn is equal to the LZ complexity.

\subsection{Gambling}

In \cite{Feder91},
sequential gambling schemes, where the amount to be wagered
on the future outcome is obtained by a finite-state machine were
analyzed. In that work, the finite-state machine calculates the
percentage of the wagered capital at time instant $i$ on the outcome at
the next time instant, $i+1$, and that wagers are paid at even odds. The
maximum capital attained by any finite-state machine is characterized in terms
of the finite-state
complexity of the given individual sequence is proved. A concrete gambling
scheme was then
proposed based on the incremental parsing process of the LZ78 algorithm.
The capital achieved was found and it turned out that asymptotically,
its exponential growth is as fast as the exponential growth
achieved by any finite-state gambling machine. 

\subsection{Prediction}

A year later, in \cite{FMG92}, the related problem of universal prediction of
binary individual sequences using finite-state predictors was addressed. 
The model adopted was in the spirit of the one in eqs.\ (\ref{output}) and
(\ref{next-state}), except that in eq.\ (\ref{output}), the output was defined
to be an estimate of the next outcome of the sequence, namely,
\begin{equation}
\hat{x}_{i+1}=f(z_i,x_i),
\end{equation}
and the performance of a predictor was measured in terms of the relative
frequency of prediction errors in the long run. A notion of {\em finite-state
predictability} was defined under the inspiration of \cite{ZL78}, as the
asymptotic minimum fraction of prediction errors attainable by any finite-state
predictor similarly as in the above mentioned definitions associated with
compressibility. A mechanism similar to the one in \cite{Feder91} was used for
universal prediction scheme that asymptotically achieves the finite-state
predictability. This was achieved by devising a running empirical
conditional probability distribution (based on the LZ phrases) of the next outcome given the past.
If the empirical conditional probability of `1' was well above $\frac{1}{2}$,
then the predictor would guess that the next out come is $\hat{x}_{i+1}=1$. If
it was significantly below $\frac{1}{2}$, the guess would be
$\hat{x}_{i+1}=0$. In the vicinity of $\frac{1}{2}$, the prediction was
randomized. The ideas of \cite{FMG92} were extended later in various
directions, as summarized (among other things) in the tutorial article \cite{MF98}.

\subsection{Filtering}

In the filtering problem considered in
\cite{OWWSM04}, a finite-alphabet individual sequence is
corrupted by a memoryless channel and the objective was to reconstruct
the underlying clean
sequence, with as low distortion as possible, by processing 
the channel output sequence causally.
Using the incremental parsing procedure, practical
filtering algorithms were devised. In particular,
a finite-memory filter of order $k$
was defined to have the
property that the estimation at any time instant is a time-invariant function
of the channel outputs from time $t-k$ to time $t$, inclusive.
The universal filter derived was shown to 
achieve distortion essentially as small as that
of the best finite--memory filter of any fixed order,
that is informed with
full knowledge of the clean sequence.
More general finite-state filters were also considered and it was shown that any such filter is
well approximated by some finite-memory filter of growing order, and so,
universality of the proposed algorithms was established with respect to this larger class.

\subsection{Guessing}

Motivated by earlier work on universal randomized guessing, in \cite{me20},
the individual-sequence
setting  was studied in the context of the guessing problem: in this setting,
the objective was to
guess a secret, individual (deterministic) vector 
$x^n= (x_1,\ldots,x_n)$, by using a finite-state machine that sequentially
generates randomized guesses from a stream of purely random bits.
The finite-state
guessing exponent was defined as the asymptotic normalized logarithm of the minimum
achievable $\rho$th order moment
of the number of randomized guesses, generated by any finite-state machine,
until $x^n$ is guessed
successfully. It was shown in \cite{me20} that the finite-state guessing exponent of any sequence
is intimately
related to its finite-state compressibility, and it is
asymptotically
achieved by the decoder of (a slightly modified version of) the LZ78
algorithm, fed by purely random
bits. The results in \cite{me20} are also
extended to the case where the guessing machine has access to a side
information sequence, $y^n=
(y_1,\ldots,y_n)$, which is also an individual sequence.

\subsection{Universal Code Ensembles}

In \cite{me23}, a universal ensemble for random selection of rate-distortion codes,
which is asymptotically optimal in an individual-sequence sense was proposed. According to this ensemble,
each reproduction
vector, $\hat{x}^n$, is selected independently at random under the universal probability
distribution, 
\begin{equation}
P_{\mbox{\tiny univ}}(\hat{x}^n)=\frac{2^{-c(\hat{x}^n)\log c(\hat{x}^n)}}{Z}
\end{equation}
where $Z$ is the normalization constant,
\begin{equation}
Z=\sum_{\hat{x}^n}2^{-c(\hat{x}^n)\log c(\hat{x}^n)},
\end{equation}
which echoes the spirit of the universal distribution defined in the context of the
Kolmogorov complexity \cite[Section 14.6]{CT06}.
It is shown that with high probability, the randomly drawn
codebook yields
an asymptotically optimal variable-rate lossy encoder with respect 
to an arbitrary
distortion measure, as a compatible converse theorem holds as well.
According to the
converse theorem, even if the decoder knew $\ell$-th order type class of
source vector ahead of time
($\ell$ being
a large but fixed positive integer), the rate-distortion performance
code could not have
been improved, for most of the codewords that represent
source sequences within
in the same type. This establishes an individual-sequence analogue of the rate
distortion function in the form of
\begin{equation}
\varrho(x^n,D)=-\frac{\log P_{\mbox{\tiny univ}}\{\calB(x^n,D)\}}{n},
\end{equation}
where $\calB(x^n,D)=\{\hat{x}^n:~d(x^n,\hat{x}^n)\le nD\}$, $d(\cdot,\cdot)$ is a
(not necessarily additive) distortion measure that satisfies certain
regularity conditions, and $D$ is the distortion level.
This rate-distortion  performance is easily seen to be better than that of the scheme that selects
the reproduction vector with the shortest
LZ78 code-length among all possible reproduction vectors within
$\calB(x^n,D)$.

\section{Summary and Outlook}

In this article, we reviewed one of the most monumental contributions of Jacob Ziv
to information theory - the individual-sequence approach. We started from 
the jewel in the crown - the LZ algorithm in its many versions. The LZ
algorithm is a special example of the rare combination of
a beautiful theory on the one hand,
and great practicality, on the other hand. Our main focus, in this article,
was on an aspect that is probably less familiar to the general Information
Theory community -- the utility of LZ compression, and in particular, the
incremental parsing procedure, across a wide spectrum of information processing tasks beyond data
compression. This broad utility indicates that there must be something very
deep associated with the ability of the incremental-parsing mechanism to
gather statistics from data in a profound sense. 
Ziv's inequality, which relates the probability of sequence to the number of
phrases, plays a pivotal role in harnessing the LZ algorithm as an engine for
the other tasks. Without any doubt,
Ziv's legacy has influenced my own research journey, as well as those of
several colleagues and
former students, and we have seen here only a small fraction of many examples
of this fact. I am sure that this legacy will continue to influence my research work for
years to come, as I am still fascinated by its beauty and elegance. 
One challenge that might be interesting to explore in the future is about
extensions to multiuser network configurations.


\begin{thebibliography}{9}

\bibitem{GRSPMLAC21}
B.~Guo, S.~Ro, A.~Shih, T.~Phan, S.~Martiniani,
D.~Levine, R.~Austin, and P.~Chaikin, ``Capturing the local entropy production
by data compression,'' {\em APS March Meeting}, 2021.

\bibitem{Chaitin66}
G.~J.~Chaitin, ``On the length of programs for computing binary sequences,''
{J.~ACM}, vol.\ 13, pp.\ 547--569, 1966.

\bibitem{CT06}
T.~M.~Cover and J.~A.~Thomas, {\em Elements of Information Theory}, Second
Edition, Wiley - Interscience, Hoboken, New Jersey, U.S.A., 2006.

\bibitem{Feder91}
M.~Feder, ``Gambling using a finite state machine,''
{\em IEEE Trans. Inform. Theory},
vol.~37, no.~5, pp.~1459--1465, September 1991.

\bibitem{FMG92}
M.\ Feder, N.\ Merhav, and M.\ Gutman,
``Universal prediction of individual sequences,''
{\em IEEE Trans. Inform. Theory},
vol.~38, no.~4, pp.~1258--1270, July 1992. 

\bibitem{Kolmogorov65}
A.~N.~Kolmogorov, ``Three approaches to the quantitative definition of
information,'' {\em Probl.\ Inform.\ Transm.\ (USSR)}, Vol.~1, pp.\ 4-7, 1965. 

\bibitem{Kolmogorov68}
A.~N.~Kolmogorov, ``Logical basis for information theory and probability theory,''
{\em IEEE Trans.~Inform.~Theory\/}, vol.\ IT-14, pp.\ 662-664, 1968.

\bibitem{LZ98}
A.~Lapidoth and J.~Ziv, ``On the universality of the LZ-based decoding
algorithm,'' {\em IEEE Trans.~Inform.~Theory\/}, vol.\ 44, no.\ 5, pp.\
1746--1755, September 1998.

\bibitem{LZ76}
A.~Lempel and J.~Ziv, ``On the complexity of finite sequences,''
{\em IEEE Trans.~Inform.~Theory\/}, vol.\ IT-22, no.\ 1, pp.\ 75-81, January 1976.

\bibitem{me00}
N.~Merhav, ``Universal detection of messages via finite--state channels,''
{\it IEEE Trans.\ Inform.\ Theory},
vol.\ 46, no.\ 6, pp.\ 2242--2246, September 2000.

\bibitem{me13}
N.~Merhav, ``Perfectly secure encryption of individual sequences,''
{\it IEEE Trans.\ Inform.\ Theory}, vol.\ 59, no.\ 3, pp.\ 1302--1310, March
2013.

\bibitem{me14}
N.~Merhav, ``On the data processing theorem in the semi--deterministic
setting,'' {\it IEEE Trans.\ Inform.\ Theory},
vol.\ 60, no.\ 10, pp.\ 6032--6040, October 2014.

\bibitem{me15}
N.~Merhav, ``Sequence complexity and work extraction,''
{\it Journal of Statistical Mechanics: Theory and Experiment},
P06037, June 2015. doi:10.1088/1742-5468/2015/06/P06037

\bibitem{me20}
N.~Merhav, ``Guessing individual sequences: generating randomized guesses
using finite--state machines,'' {\it IEEE Trans.\ Inform.\
Theory}, vol.\ 66, no.\ 5, pp.\ 2912--2920, May 2020.

\bibitem{me21}
N.~Merhav, ``Encoding individual source sequences for the wiretap channel,''
{\em Entropy}, 2021,
23(12), 1694; {\tt https://doi.org/10.3390/e23121694}
December 17, 2021.

\bibitem{me23}
N.~Merhav, ``A universal ensemble for sample-wise lossy compression,''
{\em Entropy}, 2023, 25(8), 1199;
{\tt https://doi.org/10.3390/e25081199}, August 2023.

\bibitem{MF98}
N.~Merhav and M.~Feder, ``Universal prediction,'' (invited paper)
{\it IEEE Trans.\ Inform.\
Theory}, vol.\ 44, no.\ 6, pp.\ 2124--2147, October 1998.

\bibitem{MGZ89}
N.\ Merhav, M.\ Gutman, and J.\ Ziv, ``On the estimation of the order of
a Markov chain and universal data compression,''
{\em IEEE Trans.\ Inform.\ Theory},
vol.~35, no.~5, pp.~1014--1019, September 1989.

\bibitem{MZ06}
N.~Merhav and J.~Ziv, ``On the Wyner--Ziv problem for individual sequences,''
{\it IEEE Trans.\ Inform.\ Theory}, vol.\ 52, no.\ 3, pp.\ 867--873, March
2006.

\bibitem{Oh23}
S.~Oh, ``Universal Anomaly Detection and Applications,''
M.Sc.~thesis, Electrical and Computer Engineering Department,
the University of Michigan, Ann Arbor, Michigan, U.S.A., 2023.

\bibitem{OWWSM04}
E.~Ordentlich, T.~Weissman, M.~J.~Weinberger, A.~Somekh-Baruch, and
N.~Merhav, ``Discrete universal filtering through incremental parsing,''
{\em Proc.\ 2004 Data Compression Conference (DCC 2004)}, Snowbird, UT,
U.S.A., March 2004.

\bibitem{PCF05}
D.~Pereira Coutinho and M\'ario A.~T.~Figueiredo, ``Information theoretic text
classification using the Ziv-Merhav method,'' in: Marques, J.~S., P\'erez de la
Blanca, N., Pina, P.~(eds) {\em Pattern Recognition and Image Analysis}. IbPRIA
2005. Lecture Notes in Computer Science, vol.\ 3523. Springer, Berlin,
Heidelberg. {\tt https://doi.org/10.1007/11492542\_44}

\bibitem{PCFF10}
D.~Pereira Coutinho, A.~L.~N.~Fred and M\'ario A.~T.~Figueiredo, ``One-lead
ECG-based personal identification using Ziv-Merhav cross parsing,''
{\em Proc.\ 20th International Conference on Pattern Recognition}, pp.\
3858-3861, August
23-26, Istanbul, Turkey, 2010.

\bibitem{PWZ92}
E.~Plotnik, M.~J.~Weinberger, and
J.~Ziv, ``Upper bounds on the probability of sequences emitted by finite-state
sources and on the redundancy of the Lempel-Ziv algorithm,''
{\it IEEE Trans.\ Inform.\ Theory}, vol.\ 38, no.\ 1, pp.\ 66--72, January
1992.

\bibitem{Solomonoff64}
R.~J.~Solomonoff, ``A formal theory of inductive inference,'' {\em Inform.\
Control}, vol.\ 7, no.\ 1, pp.\ 224--254, 1964.

\bibitem{UK03}
T.~Uyematsu and S.~Kuzuoka, ``Conditional Lempel-Ziv 
complexity and its
application to source coding theorem with side information,''
{\em IEICE Trans.\ Fundamentals}, Vol.\ E86-A, no.\ 10, pp.\ 2615--2617,
October 2003.

\bibitem{Zivxx}
J.~Ziv, ``Perfect secrecy for individual sequences,'' unpublished manuscript,
1977.

\bibitem{Ziv78}
J.~Ziv, ``Coding theorems for individual sequences,'' 
{\em IEEE Trans.~Inform.~Theory\/},
vol.~IT--24, no.~4, pp.~405--412, July 1978.

\bibitem{Ziv80}
J.~Ziv, ``Distortion-rate theory for individual sequences,'' 
{\em IEEE Trans.~Inform.~Theory\/},
vol.~IT--26, no.~2, pp.~137--143, March 1980.

\bibitem{Ziv84}
J. Ziv, ``Fixed-rate encoding of individual sequences with side 
information,'' {\em IEEE Transactions on Information Theory\/},
vol.~IT--30, no.~2, pp.~348--452, March  1984.

\bibitem{Ziv85}
J.~Ziv, ``Universal decoding for finite-state channels,'' 
{\em IEEE Trans.~Inform.~Theory\/},
vol.~IT--31, no.~4, pp.~453--460, July 1985.

\bibitem{Ziv88a}
J.~Ziv, ``On classification with empirically-observed statistics and 
universal data compression,''
{\em IEEE Trans.~Inform.~Theory\/},
vol.~IT--34, no.~2, pp.~278--286, March 1988.

\bibitem{Ziv88b}
J.~Ziv, ``Compression, tests for randomness, and estimating the 
statistical model of an individual sequence,''
{\it Sequences -- Combinatorics, Compression, Security, and Transmission},
R.\ M.\ Capocelli Ed., New York: Springer Verlag, pp. 366--373, 1990.

\bibitem{ZL77}
J.~Ziv and A.~Lempel, ``A universal algorithm for sequential data 
compression,'' {\em IEEE Trans.~Inform.~Theory\/},
vol.~IT--23, no.~3, pp.~337--343, May 1977.

\bibitem{ZL78}
J.~Ziv and A.~Lempel, ``Compression of individual sequences via 
variable-rate coding,''
{\em IEEE Trans.~Inform.~Theory\/},
vol.~IT--24, no.~5, pp.~530--536, September 1978.

\bibitem{ZM92}
J.\ Ziv and N.\ Merhav, ``Estimating the number of states of a finite-state
source,'' {\em IEEE Trans.\ Inform.\ Theory},
vol.~38, no.~1, pp.~61--65, January 1992.

\bibitem{ZM93}
J.\ Ziv and N.\ Merhav, ``A measure of relative entropy between individual
sequences with application to universal classification,''
{\em IEEE Trans.\ Inform.\ Theory},
vol.~39, no.~4, pp.~1270--1279, July 1993.

\end{thebibliography}
\end{document}